\documentclass[journal=jacsat,manuscript=article]{achemso}
\usepackage{graphicx}
\usepackage{dcolumn}
\usepackage{bm}

\title{Deformation potentials and electron-phonon coupling in silicon nanowires} 
\author{F.~Murphy-Armando}
\email{philip.murphy@tyndall.ie}
\author{G.~Fagas}
\author{J.~C.~Greer}
\affiliation{Tyndall National Institute, University College Cork, Lee Maltings, Cork, Ireland}
\date{\today}
\begin{document}
\begin{abstract}
The role of reduced dimensionality and of the surface on electron-phonon (e-ph) coupling in silicon nanowires is determined from first principles. Surface termination and chemistry is found to have a relatively small influence, whereas reduced dimensionality fundamentally alters the behavior of deformation potentials. As a consequence, electron coupling to `breathing modes' emerges that can not be described by conventional treatments of e-ph coupling. The consequences for physical properties such as scattering lengths and mobilities is significant: the mobilities for [110] grown wires is 6 times larger than for [100] wires, an effect that can not be predicted without the form we find for Si nanowire deformation potentials.

\end{abstract}

\maketitle

Silicon nanowires, beyond having been successfully demonstrated as conventional semiconductor devices \cite{cui2,cui4,duan,Goldberger}, are beginning to be seen as the important blocks for novel energy harvesting applications, such as solar cells~\cite{tian,Tsakalakos} and efficient thermoelectric devices~\cite{boukai,hoch}. All these applications rely on a high electronic conductivity of electrons, while in the latter a low phonon conductivity is also essential~\cite{Galli,Shi}. At the heart of understanding the interrelation between these properties is the interaction of phonons and electrons: scattering of electrons with phonons reduces the conductivity of these devices, increasing heating and reducing their efficiency, while the scattering of phonons with electrons reduces thermal conductivity, yielding better thermoelectric attributes. But although electron-phonon coupling in bulk semiconductors has been extensively studied, the nanoscale regime is not so well understood.

Materials when structured as low-dimensional systems are known to exhibit new behavior relative to their macroscopic properties. In the case of silicon nanowires, quasi-one dimensional structures, band folding can result in an electronic structure with a direct band gap. The Increased surface to volume ratio permits the surface chemistry to significantly alter the value of band gaps ~\cite{cho2,nolan}. In this paper, we demonstrate that reduced dimensionality in silicon nanowires also has a significant impact on electron-phonon (e-ph) interactions that has not been previously anticipated. We derive a new form for deformation potentials in Si nanowires that enables a general approach to the calculation of e-ph couplings, an approach that may be readily extended to other semiconductor nanowires. The validity of such macroscopic theories
at the nanoscale is central to advancing semiconductor technology (see, e.g., the developments in the effective mass approximation~\cite{EM1,EM2,EM3}). To our knowledge, this is the first paper to address the effects of dimensionality on the deformation potential theory.


For this purpose, we employ density functional theory (DFT) to calculate the band structure of Si nanowires
and apply strain to extract the deformation potentials. DFT is known to give good agreement with experiments of the band gap pressure dependence\cite{chang,zhu} and to give accurate deformation potentials that reproduce the n-type carrier mobility in bulk\cite{prb}. Focusing on electrons, we find that contrary to common
assumption, the tabulated deformation potentials not  only vary as a function of size and orientation but also vary with
respect to the direction of the applied strain. This has a direct influence on the strength of the scattering of electrons by phonons travelling in different directions. We develop a simple theory to take into account the full anisotropy of the deformation potentials and its impact on electron  mobility.
The effect of the direction of growth of the wires and surface passivation is also studied. Notably, the deformation potentials in [110]-oriented wires are found to be highly anisotropic when compared to those of [100] wires or bulk silicon. This results in the suppression of the scattering from breathing modes, and, coupled to their lower effective mass, leads to much higher mobilities for [110] wires.

Electron-phonon coupling in nanowires can be expressed in a similar manner to the 
deformation potential theory for anisotropic scattering in multi-valley semiconductors~\cite{herrvogt}. 
A first approximation might be to use the bulk values for the deformation potentials to determine an
isotropic model independent of the nanowire diameter or surface termination together with the effective masses
calculated for specific nanowires. Unfortunately, as Leu {\it et al}~\cite{cho} find in their recent paper,
the deformation potentials of the valence and conduction bands of $\left[110\right]$ and $\left[111\right]$
Si nanowires are substantially different from those in bulk Si for strains applied along the nanowire axis.
A common practice is to assume an isotropic e-ph scattering rate with an effective value to account for
deformation potential variations in nanowires relative to bulk~\cite{zou,jin,Knezevic}, 
greatly simplifying the calculation of physical quantities derived from the deformation potential.
Speaking against this approach is the fact that Fischetti and Laux~\cite{fischetti} have shown that the anisotropy of 
the e-ph matrix element has a substantial effect on the magnitude of the electron-phonon scattering rate in a 2D electron gas 
in silicon. 

The effect of an acoustic phonon on the electronic band structure is equivalent to that produced by a slowly varying strain in the direction of the phonon. This strain introduces a correspondent slowly varying potential that scatters the electrons. The change in the energy of a band per unit of strain is called a deformation potential. The deformation potential theory for bulk semiconductors has been developed by Herring and Vogt~\cite{herrvogt}, who find the anisotropic electron-acoustic phonon matrix element for an elliptical valley in silicon to be:
\begin{equation}\label{hkq}
H_{\bf q}^\zeta({\bf r})=  \left\{  
\Xi_d   \delta{\bf R}_0^\zeta \cdot {\bf q} + 
\Xi_u   (\hat{\bf k}_\alpha \cdot \delta{\bf R}_0^\zeta)   ({\bf q} \cdot \hat{\bf k}_\alpha) 
\right\}\cos({\bf q\cdot r}),
\end{equation}
where $\Xi_d$ and $\Xi_u$ are the dilatation and the uniaxial-shear deformation potentials, respectively, $\bf q$ is the phonon momentum, $\hat{\mathbf{k}}_\alpha$ is a unit vector parallel to the $\textbf{k}$-vector of valley $\alpha$ and $\delta {\bf R}_0^\zeta $ is the amplitude of the displacement of the atoms due to a phonon with polarization $\zeta$. We will initially assume that this theory 
still holds for SiNWs, and find $\Xi_d$ and $\Xi_u$ from electronic structure calculations. Our assumption
will hold for wires with large enough diameter D. For quasi-one dimensional nanowires (D $\le 3$ nm)~\cite{note1},
we show that corrections are needed. It will be seen that, for the quasi-one dimensional wires, $\Xi_d$ becomes anisotropic
with respect to the direction of the applied strain (or phonon); this effect is particularly pronounced for $[110]$-SiNWs.


The band structure of nanowires can be understood to be due to the spatial confinement of the bulk electronic bands. 
For a $\left[100\right]$ nanowire, the indirect conduction band (CB) minimum at $\pm 0.30 \frac{\pi}{a}$ is formed by the bulk Si valleys at $2\pi/a_0\left[\pm\xi,0,0\right]$, with energy $E_x$, where $\xi=0.85$. The CB minima at $\Gamma$ are composed from the four $\Delta$ 
valleys at $2\pi/a_0\left[0,\pm\xi,0\right]$ and $2\pi/a_0\left[0,0,\pm\xi\right]$, with energies $E_y$ and $E_z$, respectively. From  
explicit band calculations for a nanowire, the four valleys that fold into $\Gamma$ are not exactly degenerate owing to level splitting 
introduced by the surface. We can use a simple theory in which the effect of the surface is to introduce interactions 
between four degenerate bands, so that the CB minima in the SiNW can be described by the eigenvalues of the following matrix,
\begin{equation}
E=\left(\begin{array}{cccc}
E_y&\alpha&\beta&\beta\\
\alpha&E_{y}&\beta&\beta\\
\beta&\beta&E_z&\alpha\\
\beta&\beta&\alpha&E_{z}
\end{array}\right),
\end{equation}
where $\alpha$ and $\beta$ are the g-type and f-type matrix elements between the bulk-like bands, respectively, introduced by the presence of the surface. The four eigenvalues of this matrix are
\begin{eqnarray}\label{eig1}
E_{1,2}&=&E_{z,y}-\alpha \\
\label{eig4}E_{3,4}&=&\frac{1}{2}\left[E_z+E_y+2\alpha\pm\sqrt{16\beta^2+\left(E_z-E_y\right)^2}\right].
\end{eqnarray}
In the absence of strain, $E_x$, $E_y$ and $E_z$ are degenerate and $E_1=E_2$, which we find to hold for hydrogenated SiNWs~\cite{GFnote1}.
Similarly, in a $\left[110\right]$ nanowire, the conduction band minimum is at $\Gamma$ and formed from the bulk valleys at $2\pi/a_0\left[0,0,\pm\xi\right]$, with energy $E_z$ and the band minimum close to $\pm 0.9\frac{\pi}{a}$, is formed by the bulk valleys $2\pi/a_0\left[\pm\xi,0,0\right]$ and $2\pi/a_0\left[0,\pm\xi,0\right]$, with energies $E_x$ and $E_y$. The effective mass of the lowest conduction band in the wires is thus given by the transverse mass in bulk, while the subband separation depends on the longitudinal mass. The effective masses are directly calculated for the wires from first principles (see \ref{tableprop}). The meaning of the effective masses $m_t$ and $m_l$ for the wires is as follows: $m_t$ is the effective mass along the direction of the wire, since, from an effective mass perspective, the lowest conduction band can be considered as formed by the valleys with transverse mass along the direction of the wire. Likewise, $m_l$ is such that it gives the correct separation between the first sub-bands.

The valleys shift according to Eq. (\ref{hkq}) as strain is applied, yielding the deformation potentials for each of the NW bands. 
%
%
From these shifts and Eqns. (\ref{eig1}) and (\ref{eig4}), we can obtain $\Xi_d$ and $\Xi_u$. For example, a strain in the $[100]$ direction yields,
$\Xi_d={\left(\Xi_1+\Xi_2+\Xi_3+\Xi_4\right)}/4$
where $\Xi_i$ are the deformation potentials corresponding to the $E_i$ bands. The variation of $\alpha$ and $\beta$ due to strain is 
found to be small and is ignored in the following.
The resulting values for the wires studied in this work are shown in \ref{tableprop}.
\begin{figure}
\includegraphics[width=2.9in]{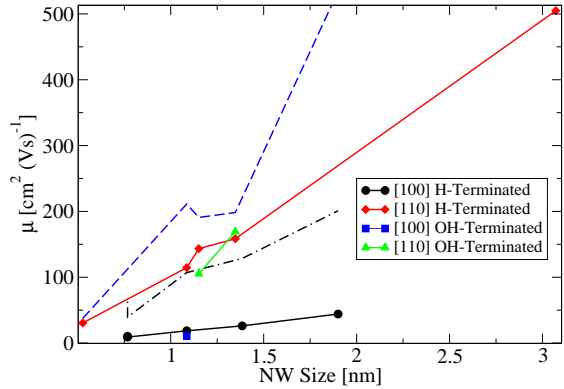}
\caption{\label{mob110100} Diameter and surface passivation dependence of the room temperature mobility of SiNWs as indicated. The dash and dot-dash lines show the the effect of disregarding the scattering by breathing modes in [110]- and [100]-oriented wires, respectively.}
\end{figure}

\begin{table*}
\caption{\label{tableprop} Calculated effective masses and splitting matrix elements for the conduction bands of the different SiNWs. Also shown are the bulk-like deformation potentials for the first CB of SiNWs.  $\Xi_{d \bot}$ and $\Xi_{d ||}$ are the $\Xi_d$ deformation potentials for phonons in the parallel and perpendicular direction to the wire, respectively (see text).The brackets next to the 
atoms per simulation cell indicate silicon crystal orientation along the nanowire axis. For the [110] wires, $\alpha$ gives the splitting of the two lowest conduction bands. $m_e$ is the electron mass.}

\begin{tabular}{ l c c c c c c c r }
\hline
\hline
Width [nm] & Atoms/cell & $m_t [m_e]$&$m_l [m_e]$&$\alpha$[eV]&$\beta$[eV]& $\Xi_{d||}$ [eV]&  $\Xi_{d \bot}$ [eV] & $\Xi_u$ [eV]\\
\hline 
0.82  & $Si_{21}H_{20}$[100]     &0.46 &1.4  &-0.3   &0.038 &-4.2&-4.2&8.2 \\
0.77  & $Si_{25}H_{20}$[100]     &0.37 &0.9  &-0.039 &0.009&-6.1&-6.1&8.7\\
1.09 & $Si_{37}H_{28}$[100]     &0.44 &1.1  &0.078  &0.025&-4.7&-4.7&8.6\\
1.26  & $Si_{57}H_{36}$[100]     &0.4  &1.1  &-0.074 &0.017&-5.5&-5.5&9.2\\
1.9   & $Si_{97}H_{44}$[100]     &0.39 &1.1  &-0.047 &0.0026&-6.6&-6.6&9.2\\
1.086  &$Si_{37}(OH)_{28}$[100] &0.72 &0.95 &0.24&0.05&-4.8&-6.5&5.3\\
0.53  & $Si_{8}H_{12}$[110]      &0.15 &1.5  &0.58   &&-9.4&-1.3&3.7 \\
1.09 & $Si_{23}H_{20}$[110]     &0.15 &0.62 &0.38   &&-10.0&-1.8&3.7\\
1.15  & $Si_{24}H_{16}$[110]     &0.15 &0.62 &0.39   &&-9.1&-0.1&3.9\\
1.35  & $Si_{42}H_{20}$[110]     &0.15 &0.62 &0.21   &&-10.5&-1.3&4.1\\
3.07  & $Si_{152}H_{40}$[110]    &0.15 &0.62 &0.018  &&-9.5&-3.3&5.8\\ 
1.15  &$Si_{24}(OH)_{16}$[110] &0.19 &0.95 &0.03   &&-7.5&-3.2&3.1\\
1.35  &$Si_{42}(OH)_{20}$[110] &0.15 &0.95 &0.33   &&-7.3&-5.1&3.9\\
\hline 
\hline
\end{tabular}
\end{table*}

We have calculated\cite{note2} the deformation potentials for the first conduction bands for wires of different sizes with 
orientations $\left[100\right]$ and $\left[110\right]$, by varying the interatomic distance by 0.1\% to 0.5\% along and perpendicular to the wire
axis. \ref{tableprop} shows the $\Xi_d$ and $\Xi_u$
calculated using the values from the lowest bands at $\Gamma$. Here, $\Xi_{d||}$ and $\Xi_{d\bot}$ are the equivalent of the $\Xi_d$ deformation potential in the parallel and perpendicular directions to the wire
axis, respectively. From these direct band structure calculations, we have found that Eq. (\ref{hkq}) still holds for small nanowires, except for hydroxylated and [110] wires, where the expression can be corrected to include the anisotropy due to the breaking of bulk symmetry, as seen below.
Surprisingly, there is very little variation in the $\Xi_d$ and $\Xi_u$ bulk-like deformation potentials with the strain direction in the $\left[100\right]$ wires, and $\Xi_u$ has the same value as bulk silicon\cite{prb}. 
On the other hand, $\Xi_d$ varies substantially with strain direction in the $\left[110\right]$ wires, and $\Xi_u$ is half its value in bulk. The anisotropy in $\Xi_d$ can be accounted for in Eq. (\ref{hkq}), by interpolating this deformation potential as a function of strain direction as follows,
\begin{equation}\label{xid}
\Xi_d^{110}(\hat{q})=q^2_x\Xi_{d||}^{110}+\left(q^2_y+q^2_z\right)\Xi_{d\bot}^{110}
\end{equation}
where the $q_i$ are the cartesian components of the dimensionless
unit vector $\hat{q}$ along the phonon momentum.  
In what follows, we use the values of \ref{tableprop} and the Boltzmann transport equation
to estimate the mobility of the SiNWs. The momentum relaxation time, ${\tau_{\bf k}}$, due to acoustic phonon scattering
can be expressed as
\begin{eqnarray}\label{tau}
\frac{1}{\tau^\pm_{\bf k}}&=&\frac{1}{2L_z L_y}
\int^1_{-1} dq \left(1-\frac{v_{k+q}}{v_k}\right) \times \nonumber \\ 
&\times & \sum_{n_z',n_y'}\sum_{m_z,m_y,\zeta}
\frac{\left |\Xi_d   {\bf \hat{e}_q}^\zeta \cdot {\bf q} + 
\Xi_u   (\hat{\bf k}_\alpha \cdot {\bf \hat{e}_q}^\zeta)   ({\bf q} \cdot \hat{\bf k}_\alpha)\right |^2}{\rho \omega^\zeta_{m_z,m_y}(q)} \times \nonumber \\
&\times &\left(n^\zeta_{m_z,m_y}(q)\pm\frac{1}{2}+\frac{1}{2}\right)  \left|G_{m_z,m_y,n_z',n_y'}\right|^2 \times  \nonumber\\
&\times &\delta(E_{k+q,n_z',n_y'}-E_{k,n_z,n_y}\pm\hbar\omega^\zeta_{m_z,m_y}(q)), 
\end{eqnarray}
where the upper and lower symbols in ``$\pm$" indicate the emission or absorption of a phonon, $L_i$ are the dimensions of a wire of rectangular cross section, with $x$ the periodic direction,
$v_k$ is the velocity of carriers with momentum $\bf k$, ${\bf \hat{e}_q}^\zeta$ the polarization of a phonon with momentum ${\bf q}$ in branch $\zeta$,  and 
\begin{eqnarray}\label{gg}
&G_{m_z,m_y,n_z',n_y'}&=\frac{4}{L_zL_y}\int_0^{L_z}\int_0^{L_y} dzdy \sin( k_zz)\sin(k_yy)\times\nonumber\\
&\times&\sin( \frac{\pi}{L_z}n_z'z)\sin( \frac{\pi}{L_y}n_y'y)e^{-i \pi( \frac{m_zz}{L_z}+\frac{m_yy}{L_y})}
\end{eqnarray}
 is the overlap factor of the wavefunctions (see Note~\citenum{note1}
 for the meaning of the indices and vectors in this context). The choice of ``effective mass" wavefunctions for Eq. (\ref{gg}) over the Kohn-Sham wavefunctions is based on ease of use and to simplify the understanding of the different contributions to the scattering from the different modes. This approximation should work for large wires. In smaller wires, the electron confinement and the matrix elements introduced by the surface push five of the six sub-bands up in energy,  beyond the range of acoustic phonon scattering at room temperature. As a result, this approximation works well for the lowest sub-band, which is the main contributor the the mobility. Some corrections should be expected for the scattering into higher sub-bands, but they are not important for the cases considered in this work. We therefore expect the main difference in comparing the different wires to come from the deformation potentials.
Eq. (\ref{tau}) can be solved numerically, and using the elastic approximation does not yield noticeable differences. The acoustic phonon-limited mobility can then be obtained as 
$
\mu(E_F)=\frac{e\left < \tau(E_F)\right >}{m_t},
$
where $m_t$ is the effective mass of a carrier in the conduction band. Thanks to the breaking of the degeneracy of the conduction bands due to the surface matrix elements and confinement observed, we expect the contribution from optical and intervalley to be small for the range of energies considered in this work.  We would like to point out that, while we have used the bulk phonon dispersions, discretised perpendicular to the wire axis, the effect of the full deformation potential remains the same, and we do not expect a large difference in the mobilities between this approach and using the phonon band structure of the wires\cite{notep}.


The room temperature mobilities $\mu(E_F=0$) of wires with varying orientation and surface termination versus wire size are summarised
in ~\ref{mob110100} where it is seen that the nanowire orientation has a much larger effect on electron mobility
than surface passivation. Due to the increased effective mass, OH surface passivation affects only the [100] growth orientation, halving the mobility compared to H-passivation for the case considered. A lower mobility is observed for the $[100]$-oriented wires compared to the $[110]$-SiNWs.
One factor yielding the lower mobilities for $[100]$ wires is their larger electron effective masses 
(see ~\ref{tableprop}). However, the electron relaxation times in the $[100]$ and $[110]$ wires differ fundamentally
in that the e-ph coupling for breathing vibrational modes in the $[110]$ nanowires are suppressed relative
to the $[100]$ wires or bulk silicon. For the $[110]$ wires, electron
scattering due to the phonon modes along the wire axis is usually more than four times as strong as that due to the
modes perpendicular to the wire axis (breathing modes), and larger than for the $[100]$ wires. In the latter, however, the 
scattering is more isotropic, with the result that breathing modes have similar e-ph coupling strengths as for excitation of 
modes along the wire axis. To demonstrate the importance of the breathing modes in suppressing the $[100]$-mobility, in \ref{mob110100} we consider the effect of de-coupling the breathing modes 
from the scattering matrix. As the nanowire diameter is increased, coupling to the breathing modes play an increasingly
important role in determining the electron mobility.

The anisotropy in the deformation potentials is also expected to influence the thermoelectric performance of NWs.
The electrical conductivity and the electronic and phonon thermal conductivity that determine the figure of merit ($ZT=\frac{\sigma S^2}{\kappa_I+\kappa_e} T$, where $\kappa_I$ and $\kappa_e$ are the ionic and electronic thermal conductivities, respectively, $\sigma$ is the electronic conductivity, $S$ the Seebeck coefficient, and $T$ is the temperature), 
will depend on several factors such as temperature, impurity concentration, surface roughness and chemistry. 
Considering the effect of the anisotropic potentials alone, better performance is expected for the [110] NWs
compared to the [100] or bulk silicon: the enhanced scattering of the phonons by electrons along the direction of
the wires (i.e. the direction of the thermal gradient) decreases thermal conductivity, while the overall reduction of the scattering of electrons by breathing modes enhances the electronic mobility. Hence, these two effects combined should result in a higher figure of merit for the [110] NW.



\begin{figure}
\includegraphics[width=3.5 in]{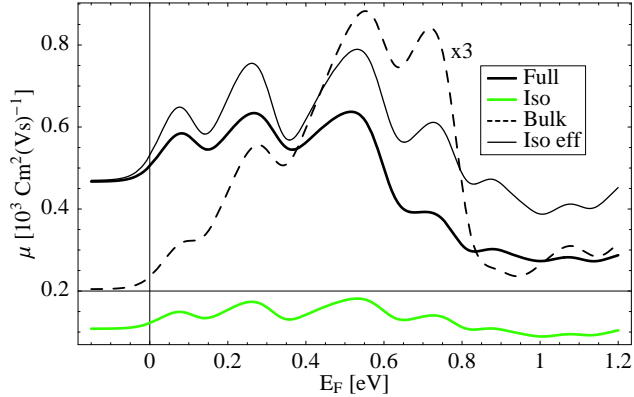}
\caption{\label{mob152-bk} (colour online)
Comparison of the room temperature mobilities vs Fermi energy of a 3nm $\left[110\right]$ wire
between isotropic models (green, thin black), bulk deformation potentials (dashed, scaled down threefold to fit the figure) and the full anisotropic electron-phonon coupling (thick black).
The bulk deformation potentials are those from Ref. \citenum{prb}, the isotropic deformation potential is that used in Ref. 
\citenum{jin}. The thin black line represents the isotropic model using the effective deformation potential that fits the anisotropic model at $E_F<<0$ (See Text). Interestingly, quantization effects corresponding to quasi-one dimensional
electron density of states persists even at room temperature. 
}
\end{figure}

We would like highlight the consequences of attempting to apply bulk-like deformation potential models to calculate the electron 
mobility in SiNWs. In ~\ref{mob152-bk}, we show the room temperature mobility versus the Fermi energy for a wire of 3 nm diameter 
oriented along a Si $[110]$ direction, computed using the deformation potentials calculated in this work and compared to the mobility 
calculated using $\Xi_d=-0.029$ eV and $\Xi_u=8.77$ eV, from Ref.
\citenum{prb} for bulk silicon. The mobility obtained by using bulk deformation potential values underestimates the scattering by longitudinal phonons along the wire direction 
resulting in a significant overestimation of the electron mobility. It has been argued\cite{jin} that it is possible to build 
an approximation by which the value of an isotropic deformation potential reproduces the universal mobility curve obtained from wires of large diameter. However, this approach underestimates the electron mobility 
relative to the use of the explicit model developed in this work, as displayed by the green curve (labeled Iso) in ~\ref{mob152-bk}.
Alternatively, to determine whether we can apply an isotropic model to reproduce the calculated mobility, we may attempt to define an effective isotropic deformation potential 
\begin{equation}\label{iso}
H_{{\rm eff} {\bf q}}^\zeta=\Xi_{\rm iso}\left({\bf\delta R}_0^\zeta  \cdot {\bf q}\right) \cos( {\bf q}\cdot{\bf r})
\end{equation}
that, used with an isotropic version of Eq.~(\ref{tau}) (with the longitudinal phonon branch only) reproduces the anisotropic mobility at $E_F\ll0$. The mobility obtained in this manner is also
plotted in ~\ref{mob152-bk}. We find that the isotropic model is still not adequate in this case, where $\Xi_d$ is highly anisotropic,
and $\Xi_u$ is smaller or near $\Xi_d$. Nevertheless, an isotropic model will work when the condition
$\Xi_d\sim\frac{3}{4}\Xi_u$ is satisfied, as is the case of the silicon nanowires oriented along a $[100]$ direction.



In summary, we have calculated from first-principles the deformation potential of nanowires of different diameters,
with $\left[100\right]$ and $\left[110\right]$ orientations and for hydrogen and hydroxyl surface termination, and
extended the deformation potential theory to develop for the first time a model that incorporates the effects of lower dimensionality in the electron phonon scattering rate. 
While we find that surface termination changes the bands substantially\cite{nolan,cho2}, it has little effect on the deformation potentials. To the contrary, we find that the deformation potentials in $\left[110\right]$ are highly anisotropic, in contrast to those of $\left[100\right]$ wires and bulk silicon. This effect results in the suppression of the scattering from breathing modes in $\left[110\right]$ wires, which, added to their lower effective mass, leads to a greater n-type carrier mobility compared to [100]-oriented wires. 

This material is based upon works supported by the Science Foundation Ireland under Grant No. 06/IN.1/I857, and the European Union project Nanosil.
We would like to thank Stephen Fahy and Baruch Feldman for useful comments on the manuscript.



\bibliography{NL_MurphyGFJG_v2_GF}

\begin{mcitethebibliography}{30}
\bibitem[GFn()]{GFnote1}
 DFT calculations were performed using the Quantum-Espresso
  package.\cite{espresso} We used the local density approximation for exchange
  and correlation, and norm conserving pseudopotentials generated by Von
  Barth-Car. A cut off energy of 408 eV for the plane-wave expansion and a
  Monkhorst-Pack $k$ point number of 4 along the NW growth direction are
  sufficient converge all paramenters within the significant figures presented.
  The interwire distance is no less than 1nm, such that there is no significant
  overlap between the wavefunctions of neighbouring wires in a supercell.
  Structures were optimized such that the force on each atom is smaller than
  $10^{-6}$ Ryd/Bohr.\relax
\mciteBstWouldAddEndPunctfalse
\mciteSetBstMidEndSepPunct{\mcitedefaultmidpunct}
{}{\mcitedefaultseppunct}\relax
\EndOfBibitem
\bibitem[not()]{note2}
 To obtain the band shifts with strain, and hence the deformation potentials,
  we subtract the band eigenvalue of the unstrained cell from that of the
  strained cell, using two different unit cell calculations. To define a
  reference potential, we ensure that the average of the local potential is the
  same for both cells in the vacuum outside the nanowire.\relax
\mciteBstWouldAddEndPunctfalse
\mciteSetBstMidEndSepPunct{\mcitedefaultmidpunct}
{}{\mcitedefaultseppunct}\relax
\EndOfBibitem
\bibitem[not()]{notep}
 We have performed calculations using the correct phonon dispersions for the
  [110] 1.15nm wire and for the [100] 1.09nm wire, and found the mobilities to
  be up to 30$\%$ higher in both cases, when compared to using the discretised
  bulk phonon dispersions. The effect of the deformation potentials remains
  unchanged, and the comparison between the two wire orientations remains
  valid.\relax
\mciteBstWouldAddEndPunctfalse
\mciteSetBstMidEndSepPunct{\mcitedefaultmidpunct}
{}{\mcitedefaultseppunct}\relax
\EndOfBibitem
\end{mcitethebibliography}
\end{document}